\documentclass[reprint,twocolumn,superscriptaddress,showpacs,preprintnumbers,amsmath,amssymb,aps, pra]{revtex4-1}

\usepackage{amssymb}
\usepackage{amsfonts}
\usepackage{amsmath}
\usepackage{cases}
\usepackage{stmaryrd}\usepackage{tipa}
\usepackage[dvips]{graphicx}
\usepackage{dcolumn}
\usepackage{bbold}
\usepackage{graphicx}
\usepackage{subfigure}
\usepackage{dcolumn}
\usepackage{bm}
\usepackage{color}
\usepackage{xcolor}
\usepackage{CJK}
\usepackage{subfigure}
\usepackage{ulem}

\newcommand{\beq}{\begin{equation}}
\newcommand{\eeq}{\end{equation}}
\newcommand{\beqa}{\begin{eqnarray}}
\newcommand{\eeqa}{\end{eqnarray}}

\begin{document}
\title{Entropies and IPR as Markers for a Phase Transition in a Two-Level Model for Atom-Diatomic Molecule Coexistence}

\author{Ignacio Baena}
\affiliation{Departamento de F\'isica At\'omica, Molecular y Nuclear, Facultad de F\'isica, Universidad de Sevilla, Apartado 1065, E-41080 Sevilla, Spain}

\author{Pedro P\'erez-Fern\'andez}
\affiliation{Dpto.\ de F\'isica Aplicada III, Escuela T\'ecnica Superior de Ingenier\'ia, Universidad de Sevilla, Sevilla, Spain }
\affiliation{Instituto Carlos I de F\'{\i}sica Te\'orica y Computacional, Universidad de Granada, Fuentenueva s/n, 18071 Granada, Spain \\}

\author{Manuela Rodríguez-Gallardo}
\affiliation{Departamento de F\'isica At\'omica, Molecular y Nuclear, Facultad de F\'isica, Universidad de Sevilla, Apartado 1065, E-41080 Sevilla, Spain}
\affiliation{Instituto Carlos I de F\'{\i}sica Te\'orica y Computacional, Universidad de Granada, Fuentenueva s/n, 18071 Granada, Spain \\}

\author{Jos\'e Miguel Arias}
\affiliation{Departamento de F\'isica At\'omica, Molecular y Nuclear, Facultad de F\'isica, Universidad de Sevilla, Apartado 1065, E-41080 Sevilla, Spain}
\affiliation{Instituto Carlos I de F\'{\i}sica Te\'orica y Computacional, Universidad de Granada, Fuentenueva s/n, 18071 Granada, Spain \\}

\begin{abstract}
A Quantum Phase Transition (QPT) in a simple model that describes the coexistence of atoms and diatomic molecules is studied. The model, that is briefly discussed, presents a second order ground state phase transition in the thermodynamic (or large particle number) limit, changing from a molecular condensate in one phase to an equilibrium of diatomic molecules-atoms in coexistence in the other one. Usual markers for this phase transition are the ground state energy and the expectation value of the number of atoms (or, alternatvely, the number of molecules) in the ground state. In this work, other markers for the QPT as the Inverse Participation Ratio (IPR) and, particularly, the Rényi entropy are analysed and proposed as QPT markers. Both magnitudes present abrupt changes at the critical point of the QPT.
\end{abstract}
%\pacs{21.60.Fw}


\date{\today}
\maketitle
\section{Introduction}
The study of phase transitions in quantum systems is a topic of present interest, usually referred to as Quantum Phase Transitions (QPT) \cite{Sachdev, Carr, Carollo19}. Since the seminal Gilmore and collaborators works \cite{Gilmore1978, Gilmore1979, Feng1981}, there have been numerous papers characterizing QPTs in two-level quantum systems of different dimensionality used to model nuclear and molecular systems, as the interacting boson model (IBM) or the vibron model (See refs.~\cite{Cejnar2009, Casten2009, Cejnar2010} and references therein).

In connection with this, particular Hamiltonians based on algebraic structures that are exactly solvable have been proposed, as the Lipkin \cite{Lipkin}, the Bose-Hubbard \cite{BH}, the Jaynes-Cummings, the Tavis-Cummings or the Dicke models \cite{Jaynes,Tavis,Dicke}, just to cite a few of them. These models present specific dynamical symmetries that correspond to different equilibrium configurations of the system in the ground state. The algebraic structure of these models allows for simple solutions in some cases what provides important references for more complex systems

In this work, a solvable two-level model that represents the coexistence of atoms and homo-nuclear diatomic molecules is used to study QPTs \cite{Tikhonenkov2008,PedroCejnar2011,NingJu2013,Graefe2015}. The model is briefly presented in Sect. II, where the matrix elements relevant in the model Hamiltonian are given explicitly in a  basis with two labels: the number of molecules and the number of atoms. The eigenvalues and eigenvectors of the Hamiltonian are easily obtained by diagonalizing the corresponding matrix. The thermodynamic or large particle number limit of the model is also presented so as to classify the QPT and analyze the critical transition point. The model has one control parameter that drives the system from a molecular condensate, in one phase, to a new phase in which atoms and molecules coexist. Usual markers for the critical point are the ground state energy and the behaviour of an order parameter that is zero in one phase and different from zero in the other one. Usually this order parameter is the expectation value in the ground state of the number of atoms (or the number of molecules). In this work, we propose the use of the Inverse Participation Ratio (IPR) and the Rényi entropy as other good markers for the critical point. 
They are presented in Sec. III. Then Sec. IV is for conclusions.
%The structure of the manuscript is the following. In Sect.2 the model is described and its exact solution is worked out. In addition its solution in the mean-field approximation is discussed an analytic expressions for the ground state energy and for the number of atoms in the system as a function of the control parameter are obtained. Sect. 3 is devoted to propose as alternative phase transition markers the IPR and different types of entropy. Finally, Sect. 4 is for summary and conclusions. 

%\medskip

\section{The model for the atom--diatomic molecule coexistence}

\begin{figure}
\includegraphics[width=0.9\linewidth]{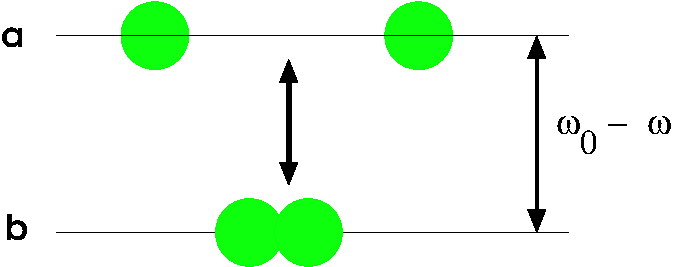}
\caption{Schematic representation of the model used in this work for the atom-diatomic coexistence. This is a two-level model. Diatomic molecules (b) are in the lower level, while single atoms (a) are in the upper level. The quantity $\omega_0-\omega$ represents the energy needed for separating the molecule into its two single atoms. This figure has been taken from \cite{PedroCejnar2011}.}
\label{f_scheme}
\end{figure}

A simple two-level model designed to describe a system of two coexisting components, individual atoms and diatomic homo-nuclear molecules, is worked out. In Fig. \ref{f_scheme}, the model is represented schematically.

Each component in the model is represented in terms of bosons. Thus, there are two-boson types: $a$ and $b$. Bosons type $a$ represent individual atoms of energy $\hbar \omega_0/2$, while $b-$bosons represent diatomic molecules of energy $\hbar \omega$. Atoms and molecules interact among them and the proposed Hamiltonian is ($\hbar=1$ is used along this work) \cite{PedroCejnar2011},

\begin{equation}
H=\frac{\omega_{0}}{2}~a^{\dagger}a+\omega~
b^{\dagger}b+\frac{\lambda}{\sqrt{2 M}}(b^{\dagger}aa+ba^{\dagger}a^{\dagger}), 
\label{Ham}\end{equation} 
where
\begin{equation}
M=2n_{b}+n_{a}
\label{M}
\end{equation}
is the total number of atoms and is a conserved quantity. This magnitude gives the size of the system. Moreover, $\hat{n}_{a}=a^\dagger a$ is the particle number operator of bosons of type $a$ (atoms) and $\hat{n}_{b}=b^\dagger b$ is the particle number operator of type $b-$bosons (diatomic molecules). The expectation value of these operators are the number of particles $n_a$ and $n_b$ of each boson type. To make everything simpler, only even M-values will be considered in this work. Also, $\lambda$ is a control parameter that drives the system from one phase to the other. 
%Due to the $M-$conservation, one can consider a basis $\{|M,~n_{b}\rangle\}$, or equivalently $\{|n_a,~n_{b}\rangle\}$, that can be used to diagonalize the Hamiltonian.
Given that $\omega_0 > \omega$, for $\lambda =0$ the ground state is just a molecular condensate without any single atom. However, as $\lambda$ increases the interaction produces a more balanced atom-molecule distribution. Thus, depending on the control parameter $\lambda$, the system presents two phases: one with just molecules and another with a molecules-atoms mixing.  

\subsection{Exact solution of the eigenvalue problem}

An obvious basis to study the Hamiltonian (\ref{Ham}) is obtained by giving the number of molecules $n_b$ and the number of individual atoms $n_a$: $|n_a,n_b\rangle$. Since $M=2n_b+n_a$ is conserved. One can use alternatively the notation $|M,n_b\rangle$ with $M$ fixed and defining the system.

The matrix elements of (\ref{Ham}) in the mentioned basis are trivial and produce a tridiagonal matrix that can be easily diagonalized for each selected $M-$value. The relevant matrix elements are:
\begin{eqnarray}
\langle M, n_b| a^\dag a |M, n_b^\prime\rangle &=& \delta_{n_b,n^\prime_b} (M-2n_b) , \\
\langle M, n_b| b^\dag b |M, n_b^\prime\rangle &=& \delta_{n_b,n^\prime_b} n_b ,  \\
\langle M, n_b| b^\dag a a|M, n_b^\prime\rangle &=& \nonumber 
\delta_{n_b,n^\prime_b+1} \sqrt{M-2n_b}\\&\times&  \sqrt{M-2n_b-1} \sqrt{n_b+1} ,\\
\langle M, n_b| b a^\dag a^\dag|M, n_b^\prime\rangle &=& \nonumber 
\delta_{n_b,n^\prime_b-1} \sqrt{M-2n_b+1}\\ &\times& \sqrt{M-2n_b+2} \sqrt{n_b} .
\end{eqnarray}

For a given $M$, the matrix to be diagonalizsed is of dimension $(M/2+1)$, since one can have from zero molecules (only $M$ atoms) to $M/2$ molecules (no atoms). A simple diagonalization of the corresponding tridiagonal matrix will provide with all Hamiltonian eigenvalues and eigenfunctions. In particular, given a $M-$number, this diagonalization allows to obtain the ground state energy and the corresponding wavefunction as a function of the control parameter $\lambda$. This can be used to study the ground state phase transition of the system as a function of $\lambda$.

In particular, once obtained the ground state wavefunction, $| gs (\lambda) \rangle$, one can use it to calculate the expectation value of the number of atoms $\langle gs (\lambda) |\hat{n}_a| gs (\lambda) \rangle$. We will show in the next subsection that this magnitude behaves as an order parameter. It is zero in one phase and different from zero in the other one. In a latter section we will show other observables that can be used as markers for the critical point of the phase transition. In order to study a reference for the phase transition, a mean field study of the model is presented now.

\subsection{Mean field for the model Hamiltonian}
In order to develop a mean field study for this model, it is useful to introduce the operators,
\begin{eqnarray}
K_+ & = & \frac{1}{2} \left( a^\dag a^\dag \right) , \\
K_- & = & \frac{1}{2} \left( a a \right) ,   \\
K_0 & = & \frac{1}{2} \left( a^\dag a + \frac{1}{2} \right) , \label{Ka}
\end{eqnarray}
that close under the $su(1,1)$ commutation relations. Using the Holstein-Primakoff expansion \cite{HP}  a new $c-$boson can be introduced as,
\begin{eqnarray}
K_+ & = & c^\dag \left(\frac{1}{2} + c^\dag c \right)^{1/2} , \\
K_- & = & \left( \frac{1}{2} + c^\dag c \right)^{1/2} c , \\
K_0 & = & \left( c^\dag c + \frac{1}{4} \right) . \label{Kc} 
\end{eqnarray}
In terms of bosons $b$ and $c$ the Hamiltonian (\ref{Ham}) can be written as,
\begin{eqnarray}
H&=&{\omega_{0}}~c^{\dagger}c+\omega~
b^{\dagger}b\\ &+& \frac{\lambda}{\sqrt{2 M}}\left[ \left( \frac{1}{2} + c^\dag c \right)^{1/2} c b^{\dagger} \right. 
+ \left. b c^\dag \left(\frac{1}{2} + c^\dag c \right)^{1/2} \right]. \nonumber
\label{Ham2}
\end{eqnarray} 
To perform a semiclassical analysis of the system, the usual relation with atom coordinates and momenta $(x,p)$ and diatomic molecule coordinates and momenta $(y,q)$ from the harmonic oscillator are introduced,
\begin{eqnarray}
\frac{c}{\sqrt{M}} &=& \frac{1}{\sqrt{2}} \left(x + i p\right) ~~~; ~~~ \frac{c^\dag}{\sqrt{M}} = \frac{1}{\sqrt{2}} \left(x - i p\right) ; \\
\frac{b}{\sqrt{M}} &=& \frac{1}{\sqrt{2}} \left(y + i q\right) ~~~; ~~~ \frac{b^\dag}{\sqrt{M}} = \frac{1}{\sqrt{2}} \left(y - i q\right) . 
\end{eqnarray}
 These are canonical transformations and in the thermodynamic limit, i.e. $M \rightarrow \infty$, the operators position and momentum commute. In addition, in this limit the factor $1/2$ can be negligible in comparison with a term multiplied by $M$. Then, introducing these relations in the Hamiltonian, it is written as,
\begin{equation}
H=H_0 + H_1 + H_2
\label{Hmf}
\end{equation}
with
\begin{eqnarray}
H_0 &=&\omega_0 \frac{M}{2} (x^2 +p^2) + \omega \frac{M}{2} (y^2 +q^2) , \\
H_1 &=&\lambda \frac{M}{2} \sqrt{(x^2 +p^2)} (x y + p q + i p y - i q x) ,  \\
H_2 &=&\lambda \frac{M}{2} (x y + p q - i p y + i q x) \sqrt{(x^2 +p^2)} .  
\label{Ham3}
\end{eqnarray} 
%%%%%%%%%%%%%%%%%%%%%%%%
To analyze the properties of the ground state of the system in the thermodynamic limit, 
it is useful to rewrite the Hamiltonian (\ref{Hmf}) using polar coordinates
$$x = r \cos \alpha ~~; ~~p = r \sin \alpha ~~ ; ~~ y = s \cos \beta ~~; ~~q = s \sin \beta. $$
Then it can be shown that the Hamiltonian (\ref{Hmf}) can be written as,
\begin{equation}
H = M [ \frac{1}{2}(\omega_0 r^2 + \omega s^2) + \lambda r^2 s \cos(\alpha - \beta)].
\end{equation}
It seems clear from this Hamiltonian that the minimum energy corresponds to $\cos (\alpha-\beta) = -1$ (it is the value that makes the second term and, therefore, $H$ minimum since the other terms are positive, $r^2$ and $s^2$). This corresponds to $\alpha- \beta=\pi$. Any choice of $\alpha$ and $\beta$ such that they differ by $\pi$ gives the minimum energy. A possible choice is $\alpha = 0$ and $\beta = - \pi$, which corresponds to $p = 0$ and $q = 0$, but any other selection of $\alpha$ and $\beta$ (and, correspondingly, of $p$ and $q$) that satisfies $\alpha- \beta = \pi$ will give the same minimum energy surface per particle 
\begin{equation}
H/M = 1/2 (\omega_0 r^2 + \omega s^2) - \lambda r^2 s
\end{equation}
\noindent which is equation (\ref{PES}) in terms of $x$ and $y$ taking  $\alpha = 0$ and $\beta = - \pi$ (or equivalently $p=q=0$), 
%both momenta, $p$ and $q$, are set to zero. Although some specific variations from zero in the value of $p$ and $q$ can produce the same minimum  energy of the system, the choice of $p$ and $q$ equals to zero ensures the study of the minimum energy surface per particle. Therefore, taking $M \rightarrow \infty$ and $p,q \rightarrow 0$, a potential energy surface is produced as a function of the coordinates $(x,y)$,
\begin{equation}
V(x,y,\lambda) = M \left[\frac{\omega_0}{2} x^2 - \frac{\omega}{2} y^2 - \lambda x^2 y \right]. 
\label{PES}
\end{equation}

This equation can be obtained in a more straightforward way from Eq. \eqref{M} using coherent boson states. However, it is interesting to illustrate some tools, as those presented above, that potentially can be used to extract finite size effects in the system (expanding the potential energy surface in 1/M powers), thus going beyond the mean field description. Anyway, Eq. (\ref{PES}) gives the classical energy surface associated to this model. 

On the other hand, the $M$ conservation leads to the condition,
\begin{equation}
 x^2 + p^2 + y^2 + q^2 = 1 ,  
\end{equation}
that for $p,q \rightarrow 0$ gives $x^2 + y^2 =1$. This allows us to reduce the original two dimensional problem to another one with only one effective degree of freedom, $x\in [-1,1]$. 
Taking into account that the sign selection $y=-\sqrt{1-x^2}$ produces lower energy than the plus sign, the energy surface per particle can be written as
\begin{equation}
v(x,\lambda)=\frac{V(x,\lambda)}{M} = \frac{\omega}{2} + \frac{\Delta \omega}{2} x^2 - \lambda x^2 \sqrt{1-x^2},
\label{Vz}
\end{equation}
where $\Delta \omega =\omega_0-\omega$.

%So, changing to polar coordinates, only one effective degree of freedom remains: $x=\cos \theta$ and $y=\sin \theta$. Consequently, the energy surface per particle can be written as 
%\begin{equation}
%v(\theta)=\frac{V(\theta)}{M} = \frac{\omega}{2} + \frac{\omega_0-\omega}{2} \cos^2 \theta + %\lambda \cos^2 \theta \sin \theta .
%\label{Vz}
%\end{equation}
The condition for minimum is
\begin{equation}
\left.\frac{d v(x,\lambda)}{dx}\right|_{x_{min}} = 0 ,
\end{equation}
%\begin{equation}
%\left.\frac{d v(\theta)}{d\theta}\right|_{\theta_c} = 0 ,
%\end{equation}
and provides two solutions,
\begin{equation}
x_{min} =  \left\{
\begin{array}{l l} 
 x_1=0 ~~~ {\rm{which~is~always~ a ~solution}} &  ;\\
  x_2 = \left(\frac{12\lambda^2-\Delta\omega^2 - \Delta \omega \sqrt{\Delta \omega^2+12 \lambda^2}}{18 \lambda^2}  \right)^{1/2} &  .
  \end{array}
\right.
\label{zetacrit}
\end{equation}
This last solution provides energy lower than $x_1=0$ when $\lambda$ is  larger than a critical value that we call  $\lambda_c$. It is also a solution of the problem the expression of $x_2$ with a plus sign in front of the square root, but the written expression, with the minus sign, gives always lower energy. The value for $\lambda_c$ can be obtained making $x_2=0$  
\begin{equation}
  12\lambda_c^2-\Delta\omega^2 \pm \Delta \omega \sqrt{\Delta \omega^2+12 \lambda_c^2} = 0,  
\end{equation}
which gives the critical point for the transition,
\begin{equation}
\lambda_c = \frac{\omega_0-\omega}{2} = \frac{\Delta \omega}{2} .
\label{lc}
\end{equation}
For given values of $\omega$ and $\omega_0$ (this fixes $\lambda_c$), the minimum energy  per particle as a function of $\lambda$ is obtained
\begin{equation}
e_0(\lambda)= \frac{E_0(\lambda)}{M} = \left\{
\begin{array}{l l}
\frac{\omega}{2} & ~~ \lambda \leq \lambda_c ,  \\
 \frac{\omega}{2} + \frac{\Delta \omega}{2} x_2^2 - \lambda x_2^2 \sqrt{1-x_2^2} & ~~ \lambda > \lambda_c. \\
\end{array}
\right.
\label{energy}
\end{equation}

\begin{figure}
\includegraphics[width=0.9\linewidth]{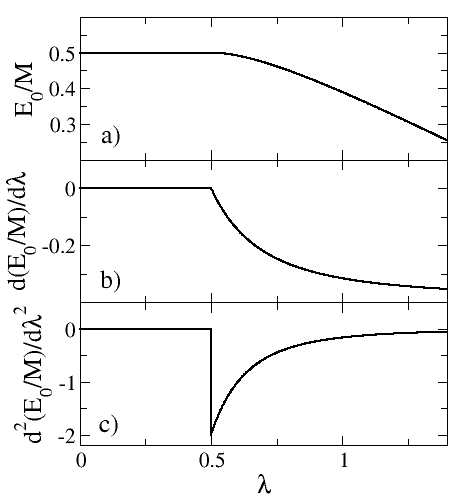}
\caption{Large-M limit (mean-field) results of the system as a function of the control parameter $\lambda$ for the case of $\omega_0=2$ and $\omega=1$. In panel a) the ground state energy per particle is represented. In panel b) its first derivative with respect to $\lambda$ is plotted. Finally, in panel c) the second derivative of the ground state energy per particle is given as a function of $\lambda$. The system undergoes a QPT for $\lambda_c=0.5$. }
%The mean field calculation is depicted in dashed-line, meanwhile the numerics for different size system are depicted in full line.}
\label{fig-gsenergy}
\end{figure}

\begin{figure}
\includegraphics[width=1.1\linewidth]{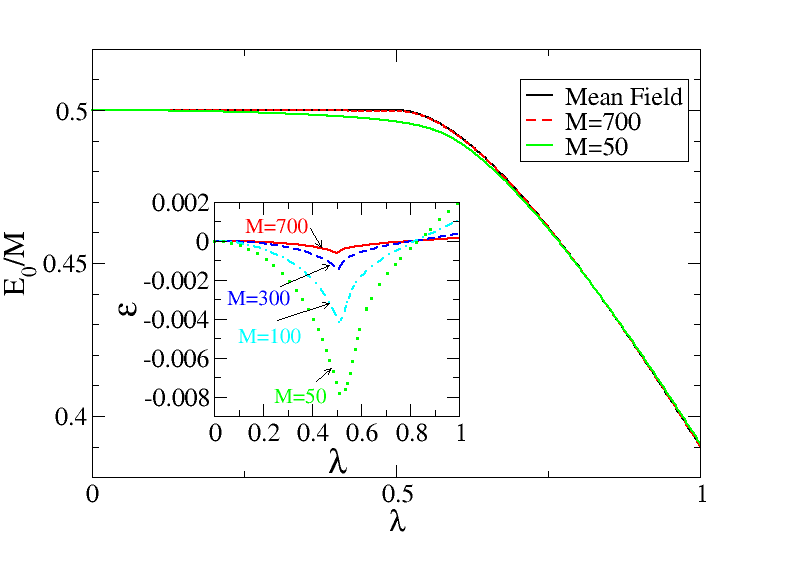}
\caption{Ground state energy per particle in the large-M limit of the system as a function of the control parameter $\lambda$ for the case of $\omega_0=2$ and $\omega=1$. The system undergoes a QPT for $\lambda_c=0.5$. The mean field calculation is depicted in black full line, meanwhile the exact numerical results for $M=50$ (full green line) and $M=700$ (dashed red line) are also presented. In order to show the convergence to the mean field with $M$, the inset represents the difference between the exact $M-$calculation and the mean field result. Different $M-$sizes are shown.}
\label{fig-gsenergy2}
\end{figure}

Eq. (\ref{energy}), with $x_2$ from Eq. (\ref{zetacrit}), gives an analytic expression for the minimum of the energy surface per particle as a function of the control parameter $\lambda$. In Fig. \ref{fig-gsenergy}, %\ref{fig-dgs} and \ref{fig-d2gs} 
the large-M limit of the ground state energy per particle (panel a), its first derivative (panel b) and its second derivative (panel c) are represented, respectively, for the case $\omega_0=2$ and $\omega=1$. In the three plots it is clear that at $\lambda=0.5$ there is a structural change in the system. This value is the $\lambda_c$ given in Eq. \eqref{lc}. Furthermore, the discontinuity of the second derivative indicates that this is a second order (or continuous) phase transition. From Fig. \ref{fig-gsenergy} it is clear that the system undergoes a second order QPT at $\lambda_c$.

Since we can solve the problem exactly for finite $M$, in Fig. \ref{fig-gsenergy2} the mean field result for the ground state energy per particle is represented, together with the exact numerical calculations with $M=50$ and $M=700$ for the case $\omega_0=2$ and $\omega=1$ that produce $\lambda_c=0.5$.
%Particularly, in Fig. \ref{fig-gsenergy} panel a), it is observed that the system undergoes a QPT for $\lambda=0.5$ according to the theoretical result. 
The mean field calculation is depicted in full black line and the exact numerical results for $M=50$ are in full green line and for $M=700$ are in dashed red line. For a size system $M=50$, the exact numerical result fits quite well to the analytical mean-field except in a small region close to the critical point (finite-size effects). Nevertheless, the bigger the size system is, the better is the agreement with the mean field calculation. This is shown in Fig. \ref{fig-gsenergy2} for $M=700$ that is basically indistinguishable from the mean-field result. In order to show better the convergence, an inset is included in Fig. \ref{fig-gsenergy2} representing a function $\varepsilon$  defined as: 
\begin{equation}
\varepsilon = \frac{E_0^{(M)}-E_0^{\rm{Mean-Field}}}{E_0^{\rm{Mean-Field}}},
\end{equation}
as a function  of $\lambda$ for different $M-$sizes.

\medskip

In addition to the energy, one can calculate analytically at the mean field level (large M limit) the expectation value for the number of atoms type $a$. From Eqs. (\ref{Ka}) and  (\ref{Kc}) one gets the relation $n_a=2 n_c$. Using the definitions of $c^\dag$ and $c$ as a function of $x$ and $p$, and taking the classical limit ($p\rightarrow 0$ and $[x,p]=0$) one obtains easily that the number of individual atoms per particle $n_a/M = x^2$. The expectation value of this observable in the system ground state is then

\begin{equation}
\langle gs |\hat{n}_a/M| gs \rangle =\langle gs | x^2| gs \rangle = \left\{
\begin{array}{l l}
0 &   ~~~~ \lambda \leq \lambda_c ,  \\
x_2^2  &  ~~~~ \lambda > \lambda_c, \\
\end{array}
\right.
\label{x2}
\end{equation}
with $x_2$ given in Eq. (\ref{zetacrit}).
This expression can be compared with real finite-M calculations to check how fast is the convergence to the large-M limit and, consequently how large are the finite-M effects.

%\begin{figure}
%\includegraphics[width=0.9\linewidth]{fig-dgs.png}
%\caption{First derivative of the ground state energy in the large-M limit of the system as a %function of the control parameter $\lambda$ for the case of $\omega_0=2$ and $\omega=1$. It %is observed that the function is continuous in all the range of values of $\lambda$.}
%\label{fig-dgs}
%\end{figure}

%\begin{figure}
%\includegraphics[width=0.9\linewidth]{fig-d2gs.png}
%\caption{Second derivative of the ground state energy in the large-M limit of the system as a %function of the control parameter $\lambda$ for the case of $\omega_0=2$ and $\omega=1$. It %is observed that the function has a discontinuity for $\lambda_c=0.5$. Therefore, the ground %state energy undergoes in $\lambda_c=0.5$ a second order (or continuous) QPT.}
%\label{fig-d2gs}
%\end{figure}

\begin{figure}
\includegraphics[width=1\linewidth]{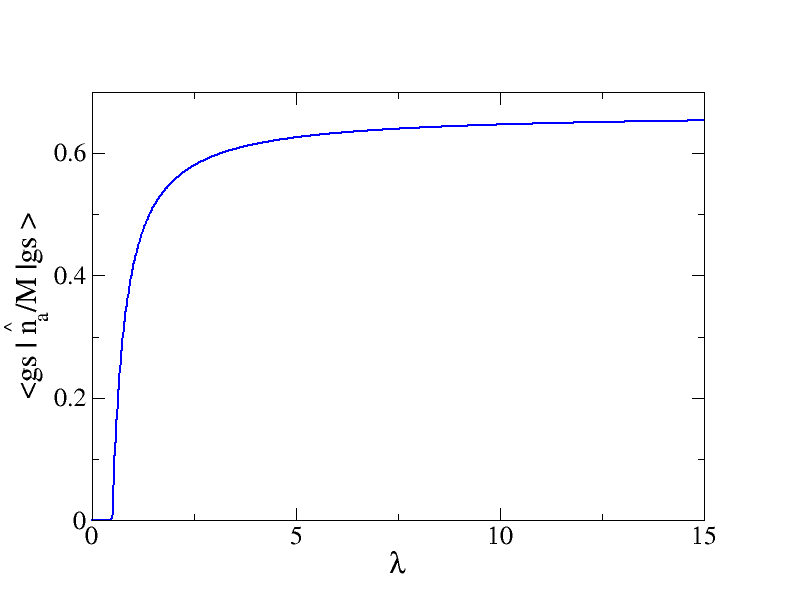}
\caption{The large-M value for $n_a/M$, number of atoms type $a$ per particle, as a function of the control parameter $\lambda$ for the case of $\omega_0=2$ and $\omega=1$. The system undergoes a QPT for $\lambda_c=0.5$. This observable behaves as an order parameter, it is zero for $\lambda<\lambda_c$ and different to zero for larger values of $\lambda$.}
\label{fig-na}
\end{figure}

\begin{figure}
\includegraphics[width=1\linewidth]{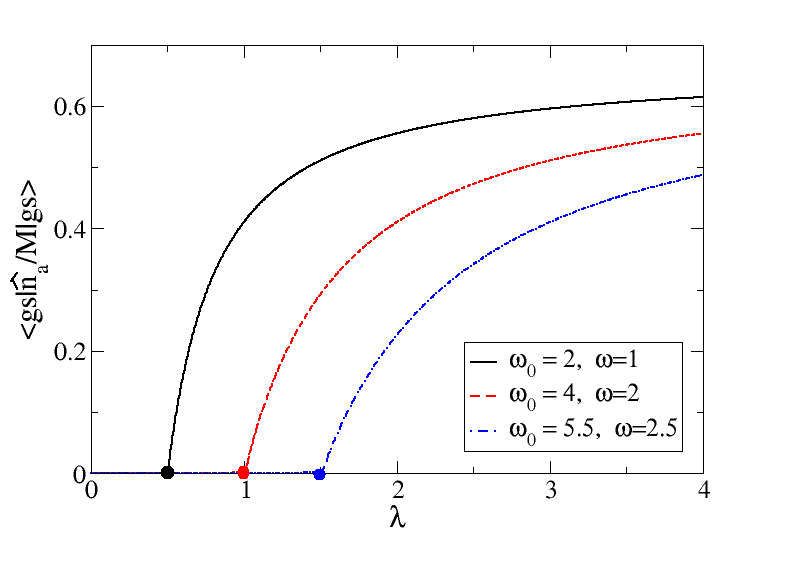}
\caption{Numerical exact calculation for the expectation value of $\hat{n}_a/M$, number of atoms type $a$ per particle, as a function of the control parameter $\lambda$ for three different cases: $\Delta \omega=1$ (full black line),  $\Delta \omega=2$ (dashed red line),  and $\Delta \omega=3$ (dot-dashed blue line). In all these cases, the critical value for $\lambda$ has been marked: $\lambda_c=0.5$, $\lambda_c=1$, and $\lambda_c=1.5$, respectively. All calculations are done for $M=700$.}
\label{fig-na2}
\end{figure}

In Fig. \ref{fig-na} the large M limit of the expectation value of $\hat{n}_a/M$ in the ground state is plotted as a function of $\lambda$. It is clear that this observable is an order parameter for the phase transition, since it is zero in one phase while different from zero in the other one. When $\lambda \rightarrow \infty$ this order parameter tends to 2/3 as given by Eqs. (\ref{x2}) and (\ref{zetacrit}). The later means means that there would be a coexistence phase of atoms and molecular particles in which it is equally likely for an atom to either be chemically bonded or to remain unbound. 
In Fig. \ref{fig-na2}, the exact calculated expectation value for $\hat{n}_a/M$ is presented for $M=700$ and three different selections for $\Delta \omega$: 1, 2, and 3, that lead to $\lambda_c$: 0.5, 1.0, and 1.5, respectively. For this large $M$ value the plots match the mean field result given by Eqs. (\ref{x2}) and (\ref{zetacrit}). It is clearly seen from Figs. \ref{fig-na} and  \ref{fig-na2} that this order parameter marks the critical point (represented in Fig. \ref{fig-na2} with filled dots for each $\omega$ selection).
%and its form is given by the analytic expression for $x^2$ given by Eqs. (\ref{x2}) and (\ref{zetacrit}).

In all cases, we have checked that the numerical results tend to the mean field approximation expressions as $M$ is increased and that the critical point corresponds to Eq. (\ref{lc}).

In our model, as in the Tavis-Cummings and Jaynes-Cummings models \cite{Larson-Irish}, quantum fluctuations are zero and, consequently, these fluctuations cannot be responsible for the corresponding vacuum instability. In this respect, some researchers consider that this is not a quantum phase transition. However, this model possesses a non-analyticity in the ground state in agreement with a continuous quantum phase transition. As such, it is a matter of taste whether the transition should be termed quantum or not.

%Let us now discuss other observables as markers for the phase transition.

\section{Other markers for the QPT}
 
 In this section we propose other markers for the critical point in the QPT.

\subsection{Inverse Participation Ratio}
The Inverse Participation Ratio (IPR) is defined as,
\begin{equation}
IPR = P^{(k)} =\frac{1}{\sum_i |c_i^{(k)}|^4}.
\label{IPR}
\end{equation}
This magnitude measures the degree of delocalization of a quantum state within a specific basis. The coefficients $c_i^{(k)}$ are the coefficients of the state $k$ in the used basis. On one hand, in case of full localization, the $k$ state is one of the basis states, then only one $c_i=1$ and the IPR will be close to 1. On the other hand, if the state $k$ is equally distributed among all basis states, then the normalization condition is
\begin{equation}
\sum_i |c_i^{(k)}|^2 = D |c_1^{(k)}|^2 =1 , 
\end{equation}
with $D$ the dimension of the matrix diagonalized. Then
\begin{equation}
  |c_i^{(k)}|= \frac{1}{\sqrt{D}}.  
\end{equation}
In this case, the maximum IPR is obtained IPR$_{max} = M/2+1$. Consequently, any values of IPR between 1 and $M/2+1$ are expected in general.

For the model discussed here, an IPR$=1$ is expected for $\lambda=0$ since in this case our Hamiltonian eigenstates are those of the harmonic oscillator. For other $\lambda-$values the Hamiltonian eigenstates will be a mixture of harmonic oscillator states and the IPR will increase. % Añadido comentario IPR 
However, not every state of the harmonic oscillator will “participate” to the eigenstate of the coexistence phase. Only a linear combination of states in which the expected number of atoms is 2/3 of M will contribute. Thus, IPR will reach a constant but smaller than the maximum possible value. 

\begin{figure}
\includegraphics[width=1\linewidth]{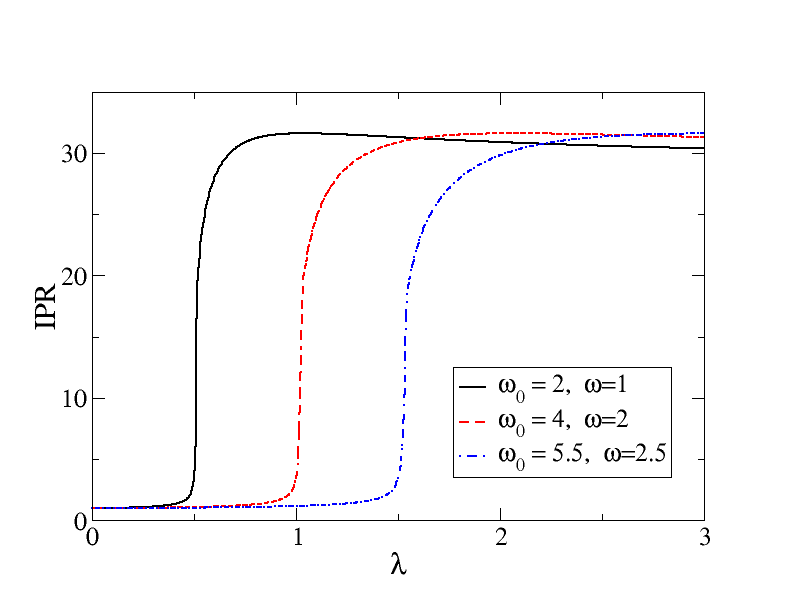}
\caption{IPR for the ground state as a function of the control parameter $\lambda$ for $M=700$ and for three different $\omega$ selections: $\Delta \omega=1$ (full black line), $\Delta \omega=2$ (dashed red line) and $\Delta \omega=3$ (dot-dashed blue line).}
\label{fig-IPR}
\end{figure}

The numerical results from the exact diagonalization of the system Hamiltonian have already been presented and these were compared to the mean-field results in the preceding section. For a given $M$, the exact diagonalization produces the ground state and, consequently, provides the coefficients $c_{M,n_b}^{\rm{gs}}$. With these, one can calculate the IPR (\ref{IPR}).
%In all cases, the ground state energy and the expectation value for the number of atoms per particle in the ground state are consistent with the mean field calculations and tend to them as $M$ is increased.
%Concerning IPR, 
In Fig. \ref{fig-IPR} the ground state IPR values  for $M=700$ and for different $\Delta \omega$ choices as a function of the control parameter $\lambda$ are presented.
The IPR marks clearly the transition of the system at the corresponding $\lambda_c$. The ground state is well localized (small IPR) in the harmonic oscillator basis for $\lambda$ below to the QPT critical point, whereas it tends to be delocalised for values of $\lambda$ above the critical value. Indeed, an abrupt change of IPR occurs at $\lambda_c=\Delta\omega/2$ in the QPT.
%%%%%%%%%%%%%%%%%%%%%%%%%%%%%%%%%%%%%%%%%%%%%%%

\begin{figure}[h!]
\includegraphics[width=1\linewidth]{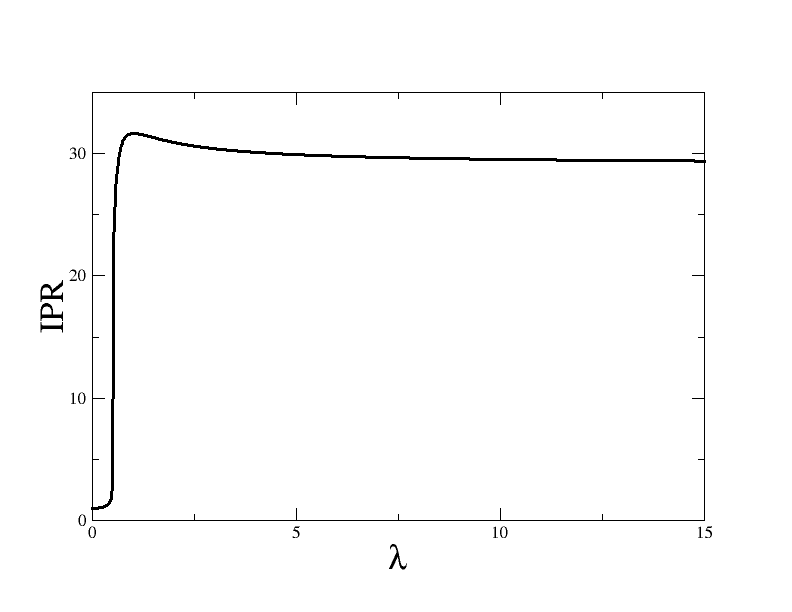}
\caption{IPR for the ground state as a function of the control parameter $\lambda$ for  $\Delta \omega =1$ and $M=700$.}
\label{fig-IPR large L}
\end{figure}

A natural question in relation to Fig. \ref{fig-IPR} is what is the asymptotic value for $\lambda \to 0$ and $\lambda \to \infty$? In order to show the $\lambda \to \infty$ we plot in Fig. \ref{fig-IPR large L} the IPR for the case $\Delta \omega =1$ and $M=700$. It is seen that the IPR for large $\lambda$ is around 30. 

Whilst an IPR= 1, or approximately 1, is expected for $\lambda \leq \lambda_c$ for which the ground state is close to a molecular condensate (the state is basically $|M, n_a=0,n_b=M/2 \rangle$), for larger $\lambda-$values the Hamiltonian eigenstates will be a mixture of harmonic oscillator states, in which there will be more than only one relevant state and, therefore, the IPR will increase. The limit of M/2+1 is obtained when all the basis states are contributing with equal weight. However, this is not reasonable and states in which the number of atoms is $n_a=2M/3$, and consequently $n_b=M/6$ (we notice that $n_a+2n_b=M$)  are expected to have a larger weight. In fact, if one assumes for the wavefunction coefficients a binomial distribution with D=M/2 components, $|M, n_a, n_b\rangle$, whose probability of $n_b$ is $p=1/3$, the corresponding IPR would be around 31. Although the distribution in our ground state is not exactly binomial, something similar is expected. In that case, the IPR will not reach the maximum possible value, and an IPR around 31 is expected for M=700.

%In order to show what $n_b$ value is expected in the large $\lambda$ limit we can take the model Hamiltonian that in the large $\lambda$ limit is 
%$$H= - \lambda (b^\dag aa + a^\dag a^\dag b).$$
%\noindent Taking the classical limit (large M) we can obtain the classical energy substituting $b$ or $b^\dag$ by $\sqrt{n_b}$ and $a$ or $a^\dag$ by $\sqrt{n_a}$. Thus, the classical energy is
%\begin{equation}
%  E = - 2 \lambda n_a \sqrt{n_b} = - 2 \lambda \left( M n_b^{1/2}- 2 n_b^{3/2}\right)
%\end{equation}
%\noindent whose minimum value corresponds to 
%$$dE/dn_b = 0  = (1/2) M n_b^{-1/2} - 3 n_b^{1/2}   \Rightarrow ~ n_b = M/6=D/3.$$

\begin{figure}[h!]
\includegraphics[width=1\linewidth]{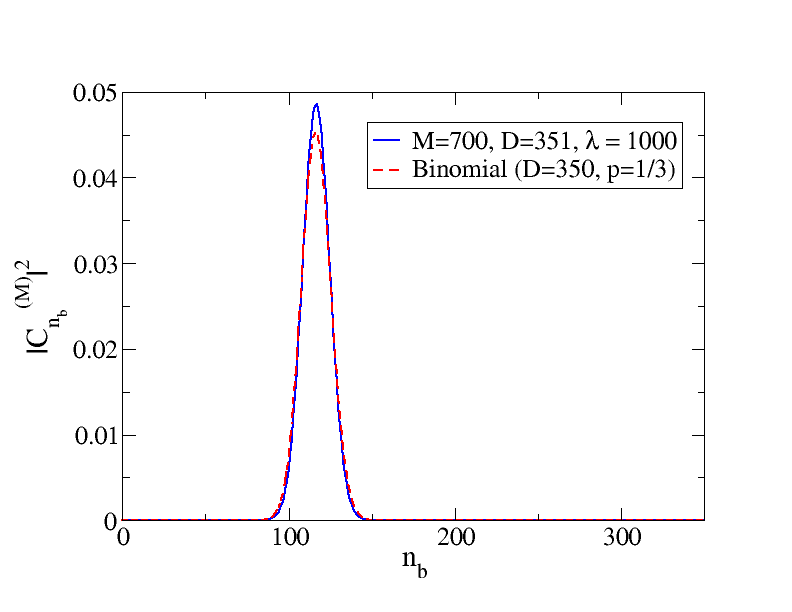}
\caption{Representation of the components of the binomial distribution for $D=350$ and $p=1/3$ (dashed red line) compared with the computed squared coefficients for the components of the ground state wavefunction in the basis $|M,n_b\rangle$ (full blue line). This last calculation was done for  $\Delta \omega =1$ with $M=700$ and $\lambda=1000$.}
\label{fig-binomial}
\end{figure}

In Fig. \ref{fig-binomial} the binomial distribution for D=350 that corresponds to M=700 (basis dimension 351) and p=1/3 (which corresponds to $n_b=M/6=D/3$) is represented  vs $n_b$ (dashed red line). Superimposed is the plot for the calculated ground state wavefunction components squared for the case M=700 and $\lambda=1000$ (full blue line). It is clearly observed the similarity of the distributions and that is why the IPR value for the large $\lambda$ limit is close to the corresponding binomial distribution (around 30 in the case of Fig. (\ref{fig-IPR large L})).  

\begin{figure}[h!]
\includegraphics[width=1\linewidth]{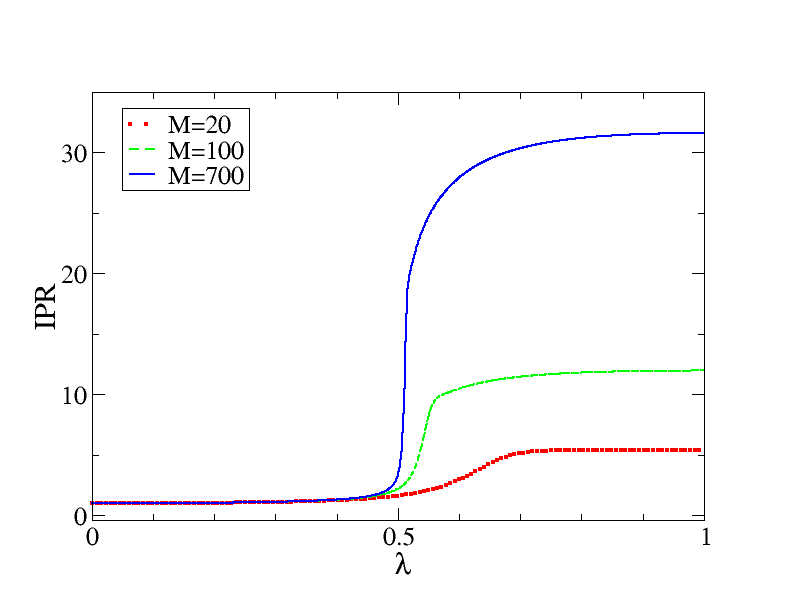}
\caption{IPR for different M-values as a function of $\lambda$. These calculations were done for  $\Delta \omega =1$.}
\label{fig-IPR-M}
\end{figure}

In order to show the behaviour of the IPR as a function of the system size, we present in Fig. \ref{fig-IPR-M} the IPR for different M-values. From this figure, we can observe that the bigger the size of the system is, the sharper is the change in the value of the IPR at the critical point.

Just a final comment on the IPR maximum observed right after the critical point. This is seen in Fig. \ref{fig-IPR large L}. This exact same behaviour is confirmed to exist for all sizes. It is not more or less accentuated depending on M. We have already established which states are relevant in both the molecular condensate phase and the coexistence phase. However, right after the critical value is reached, the state of minimum energy is given by a linear combination of a set of states wider than the one observed for large $\lambda$ values. It is a sort of transition region in which more components (fluctuations) are participating in the ground state wavefunction. As a consequence, the greatest value of the IPR is observed right there. 
%%%%%%%%%%%%%%%%%%%%%%%%%%%%%%%%%%%%%%%%%%%%%%%%%%

%marked in Fig. \ref{fig-IPR} with dots for the different $\Delta \omega$ selections.

% Añado algunas cuestiones preguntadas por el tercer revisor

% Let´s observe that the IPR seems to decrease after the critical point. This fluctuation in the IPR corresponds to a semicritical behaviour. Right after the critical value is reached, the state of minimum energy is given by a linear combination of the states relevant in both the molecular condensate phase and the coexistence phase. As a consequence, the greater value of the IPR (or maximum entropy later on) is found as observed in Fig. \ref{fig-IPR large L}. For large values of $\lambda$, the system reaches the described proportion of atoms and molecules and the IPR then becomes constant. 

\subsection{Renyi entropy}

Information was first defined rigorously by Claude Shannon \cite{Shanon}. It is a magnitude that measures how much communication it takes to transmit a message. If one has a discrete list of possible messages (events) with different probabilities, that wishes to transmit, the information value of every message depends on that probability. For instance, if one were to repeat the same message over and over, the information transmitted is measured with lower units of information. Conversely, if within this list of repeated messages something different is suddenly communicated only once, it is considered to give much more information. In other words, information measures how surprising, how unlikely, an event is. Thus, information theory does not account for content or usefulness, rather it measures only the quantity of information. The later is measured by a magnitude called entropy.

%Information theory does not account for content or usefulness, rather it measures only the quantity of information. The later is measured by a magnitude called entropy and 
%In information theory, quantum entropy 
%is a measure of the disorder and uncertainty in a system.  The entropy, \cite{QEntropy} of a state describing a physical system is a quantity expressing the diversity, uncertainty or randomness of that system. Shannon viewed this uncertainty attached to the system as the amount of information carried by its state.

Different entropies can be defined. The most popular entropy was defined by Shannon \cite{Shanon},
\begin{equation}
    S = - \int dQ ~ \rho(Q) ~\ln{\rho(Q)} ,
\end{equation}
where $Q$ are the generalised coordinates ($x$ and $y$ for our model), and  $\rho(Q)$ is the probability density ($|\Phi(x,y)|^2$, in our case). Then,
\begin{equation}
    S = - \int dx \int dy ~ |\Phi(x,y)|^2 ~\ln{ |\Phi(x,y)|^2} .
\end{equation}

%\begin{figure}[t]
%\includegraphics[width=1.\linewidth]{A_Fig-entropy.png}
%\caption{Different entropies evaluated in the ground state of the system as a function of the control parameter $\lambda$  for the case of $\Delta \omega=1$ and $M=100$. In this case, $\lambda_c=0.5$. In panel a) the Shannon entropy is plotted, in panel b) the Renyi entropy for $\alpha=1/2$ is presented, and in panel c) the Renyi entropy for $\alpha=2$ is displayed.}
%\label{fig-entropy}
%\end{figure}

Here we propose to use the Rényi entropy \cite{Renyi,Calixto} that depends on one parameter $\alpha$, for characterising the phase transition in our system. The Rényi entropy has as a limit situation the Shannon entropy ($\alpha \to 1$). The Rényi entropy is defined as,
\begin{equation}
    R^{(\alpha)} = \frac{1}{1-\alpha} \ln \left(\int dQ ~ \rho^\alpha(Q) \right), ~~ \forall \alpha \in [0,1) \cup (1,\infty)]
\end{equation}
that for the model discussed is,
\begin{eqnarray}
    R^{(\alpha)} &=& \frac{1}{1-\alpha} \ln \left(\int dx \int dy ~ |\Phi(x,y)|^{2\alpha}\right) , \nonumber \\ &~&  ~~~~~~~~~~~~~~~~~~~~~~~~ \forall \alpha \in [0,1) \cup (1,\infty)] .
\end{eqnarray}
For the model under study, the ground state is a combination of harmonic oscillator states in the coordinates $(x,y)$,
\begin{eqnarray}
  \Phi_{gs}(x,y) &=& \sum_{n_a} \sum_{n_b}  ~ c_{n_a,n_b} ~ {\cal{N}}_a H_{n_a}(x) \exp[-x^2/2] \nonumber \\ &\times& {\cal{N}}_b H_{n_b}(y) \exp[-y^2/2] ,
\label{entropy-HO}
\end{eqnarray}
where $\cal{N}$ are normalization constants and $H_n$ are Hermite polynomials. The ground state coefficients $c_{n_a,n_b}$ are obtained from the Hamiltonian diagonalization. Consequently, the entropies can be calculated with Eq. (\ref{entropy-HO}). However, this is computationally inefficient since for large M values it implies factorials of large numbers and make the calculation very heavy and inaccurate. Because of that, we prefer to go to Shannon's original idea.  The entropy \cite{Shanon} of a state describing a physical system is a quantity expressing the diversity, uncertainty or randomness of the system. Shannon viewed this uncertainty attached to the system as the amount of information carried by its state. His idea was based on the following consideration. If a physical system has a large uncertainty and one receives information on the system, then so-obtained information is more valuable (because it is less likely) than received from a system having less uncertainty. This is why entropy is measured in units of information. Shannon also drafted in \textit{ A mathematical theory of communication} \cite{Shanon} what is one of the most popular definitions of entropy. Let ${n_b}$ be a discrete random variable with probability distribution $\{ p_i\}$ of $N$ elements. That is 
\begin{equation} \notag
    \sum_{i=1}^N p_i = 1,
\end{equation}
then Shannon entropy is given by
\begin{equation}
    S=-{\sum_{i=1}^N p_i \log{p_i}}.
\end{equation}

When one takes the binary logarithm, entropy is expressed in shannons (Sh), also known as bits. Moreover, when taking the natural logarithm, as we do in this work, entropy is expressed in the natural unit of information or nat. It is merely a difference in scale (1 Sh $\approx 0.693 $ nat). Note that if one event is much more likely than the others, that is $p_j \to 1$ and $p_{i \neq j} \to 0$, then entropy tends to 0. In the opposite case, if all events were equally likely, then $p_i=1/N, \ \forall i$ and $S=\log N$, which is a function that increases with $N$. Also take notice of the fact that both the maximum and minimum possible values of Shannon entropy correspond to maximum and minimum values of IPR.

\bigskip
Should one require to measure the information provided by events giving it greater or lesser difference between likely and unlikely ones, a different definition of entropy would have to be used. 

A generalisation of Shannon entropy was made by Alfred Rényi \cite{Renyi}. Classical Rényi entropy for a parameter $\alpha \geq 0$ and $\alpha \neq 1$ is defined for the same discrete random variable as 
\begin{equation} \label{eq: classical entropy}
     R^{(\alpha)}=\frac{1}{1-\alpha} \log{\sum_{i=1}^N p_i^\alpha}. 
\end{equation}

The same minimum and maximum possible values of entropy Rényi are reached, independently of $\alpha$. In fact, the limiting value of Rényi entropy as $\alpha \to 1$, that can be calculated using L'Hôpital's rule, is the Shannon entropy $S=\lim_{\alpha \to 1} R^{(\alpha)} $. 

% Now, the quantum Rényi entropy is defined as 
% \begin{equation}
%      H_{\alpha}=\frac{1}{1-\alpha} \log{Tr(\rho^\alpha)}.
% \end{equation}

In the context of quantum theory of information, for a density matrix in a Hilbert space, $\rho \in \mathcal{N} (\mathcal{H}) $, we can define quantum Rényi entropy \cite{QuantumEntropydef} as
\begin{equation}
     R^{(\alpha)}=\frac{1}{1-\alpha} \log{Tr(\rho^\alpha)}.
\end{equation}

If $\{ p_i\}$ are the diagonal elements of $\rho$ in the basis of eigenfunctions, then the quantum Rényi entropy reduces to a Rényi entropy of a random variable $n_b$ as defined in (\ref{eq: classical entropy}). This means that for the ground state of our system we can define the probabilities $p_i=|c_i|^2$ where $c_i$ are the coefficients of the ground state wavefunction. Note that we already took the dimension of the Hamiltonian matrix $N$ as the number of elements in the discrete random distribution.  

\bigskip

For $\alpha < 1$, all random events are weighted more equally resulting in a smaller change in entropy from one state to another. %as well as a smaller value of entropy.
As $\alpha$ tends to zero, the entropy is just the logarithm of the size of the support of $n_b$, no matter the phase. 

For $\alpha > 1$, all random events are weighted more differently. As $\alpha$ grows, more likely events make larger contributions to entropy whereas less likely events are disregarded. This tends to give bigger differences between quantum phases. 

For $\alpha \to 1$, we have Shannon entropy, which results in something in the middle of both cases. 

Besides $\alpha$ dependency, there is another factor which is going to affect entropy values. As it happens with thermodynamic entropy, quantum entropy is an extensive property, meaning that it scales with the size of the system. This behaviour has already been hinted by substituting in Shannon entropy a set of values equally likely. 

On account of the above, results will be expressed according to the following criteria.

\begin{itemize}
    \item 
\textbf{Entropy dependency with $M$}
\end{itemize}
% \paragraph{Entropy dependency with $M$}

All calculations have been done for $\Delta \omega=1$ that gives $\lambda_c=0.5$.

\begin{figure}[h]
    \centering
    \includegraphics[width=1\linewidth]{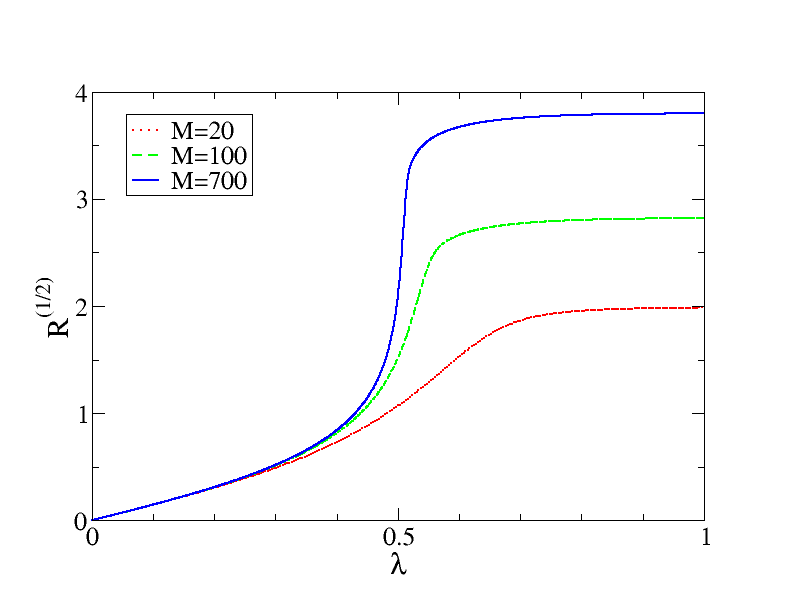}
    \caption{Rényi entropy with $\alpha=1/2$, $R^{(1/2)}$, as a function of $\lambda$ for different $M$ values.}
    \label{fig: Renyi 1/2 diff M}
\end{figure}

\begin{figure}[h]
    \centering
    \includegraphics[width=1\linewidth]{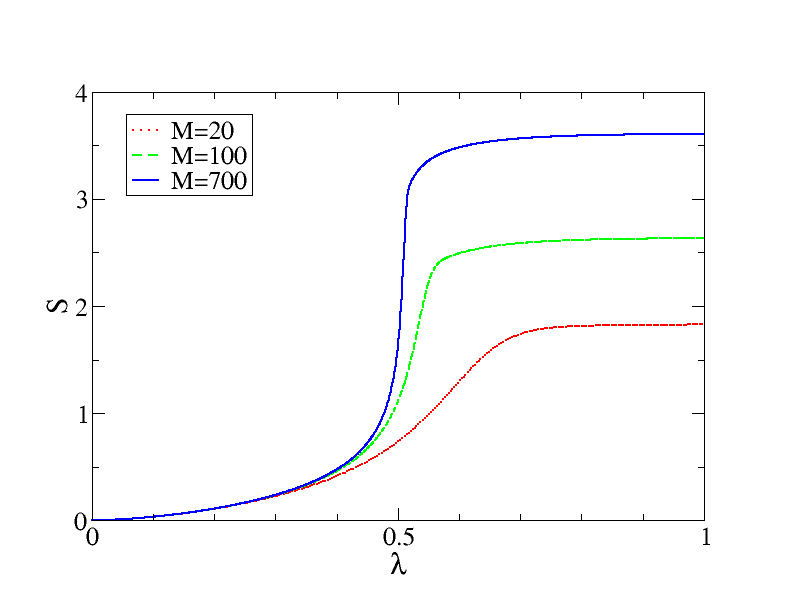}
    \caption{Shannon entropy as a function of $\lambda$ for different $M$ values.}
    \label{fig: Shannon diff M}
\end{figure}

\begin{figure}[t]
    \centering
    \includegraphics[width=1\linewidth]{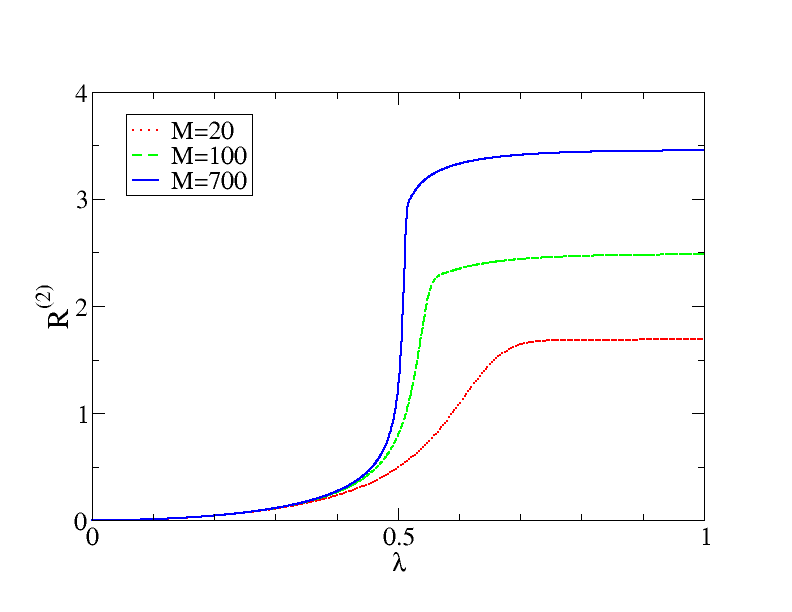}
    \caption{Rényi entropy with $\alpha=2$, $R^{(2)}$, as a function of $\lambda$ for different $M$ values.}
    \label{fig: Renyi 2 diff M}
\end{figure}

In figures \ref{fig: Renyi 1/2 diff M}, \ref{fig: Shannon diff M} and \ref{fig: Renyi 2 diff M} we can observe the dependency of different entropies with M. The transition is sharper with increasing M for all values of $\alpha$. Note that entropy is independent of $M$ in one phase but is increasingly different with larger $M-$values in the other phase. The reason behind this phenomenon lies in the characteristics of both phases. 

In the first one, the possibility of measuring the lowest eigenvalue of the harmonic  oscillator ($p_0=|c_0|^2$), is almost 1 and the rest are almost zero (which is why the IPR is approximately 1). Since the dimension $N$ is irrelevant (to a certain point), because it would not really matter how many $p_i$ there are, entropy values will be very similar and will mostly depend on $\alpha$. In the second phase, we need to reach a certain proportion of particles, given by a number of relevant coefficients that is proportional to $M$. This is the reason why the IPR also increases with $M$. 

\vspace{1cm}

\begin{itemize}
\item \textbf{{Entropy dependence with $\alpha$}} 
\end{itemize}

\begin{figure}[t]
    \centering
    \includegraphics[width=1\linewidth]{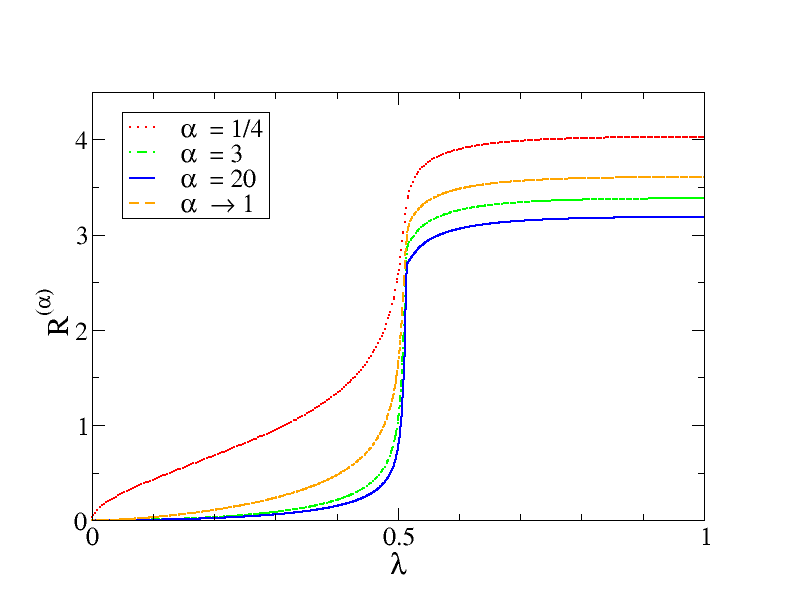}
    \caption{Different Rényi entropy values as a function of $\lambda $ for different $\alpha$ values, $M=700$ and $\Delta \omega=1$ in all cases. The case $\alpha \to 1$ is the Shannon entropy.}
    \label{fig: Entropy diff alpha M=700}
\end{figure}

In Fig. \ref{fig: Entropy diff alpha M=700}, for a system of $M=700$, the values of $R^{\alpha}(\lambda)$ are represented for a set of $\alpha$ values both under and over the unit as well as the Shannon entropy, which is given by the limit $\alpha \to 1$. It is confirmed that bigger values of $\alpha$ make the difference between both phases more evident since it distinguishes more abruptly between likely and unlikely events. It can be also confirmed that Shannon entropy is indeed between the entropy for $\alpha < 1$ and $\alpha > 1$. 

Perhaps plenty more examples and evaluations could be made toying with different values of $M$ and $\alpha$. However, the most important conclusion one would have to make is the following. The entropy, when set to an adequate $\alpha$ for it to be a good marker, is yet another quantum magnitude that experiences an abrupt (but continuous, as seen for lower $\alpha$ values) change from one phase to another, evincing the existence of a second order QPT.

\medskip

\section{Conclusions}

We have studied a two-level model for the coexistence of atoms and diatomic molecules. This model has been studied using mean-field techniques and shows a ground state second order quantum phase transition. The critical point has been obtained for the large $M-$number limit and analytic expressions for the ground state energy per particle and for the number of atoms per particle, as a function of the control parameter $\lambda$, have been worked out. This last observable is shown to be a good order parameter. We have proposed as additional markers for the phase transition the Inverse Participation Ratio (IPR) and different types of entropies. Both observables mark clearly the critical phase transition point.

\medskip

\section*{Acknowledgements}
We thank J. Gómez-Camacho for discussions.
This work is part of the I+D+i projects with Refs. PID2019-104002GB-C22 and PID2020-114687GB-I00 funded 
by MCIN/AEI/10.13039/501100011033. This work is also part of the grant Group FQM-160, EU FEDER funds US-1380840 and the  project PAIDI 2020 with Ref. P20\_01247, funded by the Consejer\'{\i}a de Econom\'{\i}a, Conocimiento, Empresas y Universidad, Junta de Andaluc\'{\i}a (Spain) and by “ERDF A way of making Europe”.

\end{document}